\documentclass[aps,twocolumn,nofootinbib,showpacs,superscriptaddress,pra]{revtex4-1}

\usepackage{graphics}
\usepackage{epstopdf}
\usepackage{bm}
\usepackage{amssymb}
\usepackage{epstopdf}
\usepackage{graphicx}
\usepackage[caption=false]{subfig}
\usepackage[cmex10]{amsmath}
\usepackage{amsthm}
\usepackage{color}
\usepackage{hyperref}
\usepackage{lipsum}

\newcommand{\ket}[1]{\left|#1\right\rangle}

\newcommand{\im}{{\rm i}}

\begin{document}

\title{Hong-Ou-Mandel interference of entangled Hermite-Gauss modes}

\author{Yingwen \surname{Zhang}}
\affiliation{CSIR National Laser Centre, PO Box 395, Pretoria 0001, South Africa}

\author{Filippus S. \surname{Roux}}
\affiliation{CSIR National Laser Centre, PO Box 395, Pretoria 0001, South Africa}
\affiliation{School of Physics, University of Witwatersrand, Johannesburg 2000, South Africa}

\author{Ebrahim \surname{Karimi}}
\affiliation{Department of Physics, University of Ottawa, 25 Templeton St., Ottawa, Ontario, K1N 6N5 Canada}
\affiliation{Department of Physics, Institute for Advanced Studies in Basic Sciences (IASBS), Zanjan 45137-66731, Iran}

\author{Andrew \surname{Forbes}}
\affiliation{CSIR National Laser Centre, PO Box 395, Pretoria 0001, South Africa}
\affiliation{School of Physics, University of Witwatersrand, Johannesburg 2000, South Africa}

\begin{abstract}
Hong-Ou-Mandel (HOM) interference is demonstrated experimentally for entangled photon pairs in the Hermite-Gauss (HG) basis. We use two Dove prisms in one of the paths of the photons to manipulate the entangled quantum state that enters the HOM interferometer. It is demonstrated that, when entangled photon pairs are in a symmetric Bell state in the Laguerre-Gauss (LG) basis, then they will remain symmetric after decomposing them into the HG basis, thereby resulting in no coincidence events after the HOM interference. On the other hand, if the photon pairs are in an antisymmetric Bell state in the LG basis, then they will also be antisymmetric in the HG basis, thereby producing only coincidence events as a result of the HOM interference.
\end{abstract}

\maketitle

\section{Introduction}

An increase in the dimensionality of quantum systems has many benefits; for example, it gives higher security in communication systems and increases the transmission rates \cite{Pasquinucci2000,Cerf2002,Bruss2002}.  There are many routes to realise high-dimensional quantum states, using various degrees of freedom, such as: time-energy \cite{Thew2004}, paths \cite{Hale2005}, the use of spatial modes \cite{zeil1, Romero2012jo, McLaren2012, Krenn2013} and combinations thereof \cite{Kwiat1998,Barreiro2005}.  In particular the entanglement of photons in higher-order Gaussian modes has generated much interest in recent years, the Laguerre-Gauss (LG) \cite{torres2,arnold2008} and Hermite-Gauss (HG) \cite{Walborn2005} modes are two of the most studied cases.

The Hong-Ou-Mandel (HOM) interference effect was first demonstrated in 1987 \cite{Hong1987}. Since then, it has become a fundamental component in many quantum information protocols and one of the defining features of quantum physics. It shows the interference of two identical photons(bosons) at a 50:50 beamsplitter where the chances of observing coincidence counts in the output ports depends on the symmetry of the two-photon state. HOM interference has been investigated for many scenarios, for example, in polarization \cite{Kwiat1992}, path length \cite{Pittman1996}, the radial dependence of the light field in a transversal plane \cite{karimi2014}, as well as in spectral filtering \cite{Giovannini2014}, for different single-photon sources \cite{Kaltenbaek2006} and with spatial modes \cite{Walborn2003,Nagali2009,Exter2007,Exter2010}. Generalisations to higher photon number and multiports, as well as realisations with other bosonic systems \cite{Luis-Swan2014} have been discussed in the literature. The HOM effect is applied to characterize single-photon sources \cite{Kock2010} but also in the construction of quantum logic gates \cite{Kok2007}, quantum cloning \cite{Ricci2004, Irving2004, NagaliNat2009, Nagali2010} and phase shaping of single photons \cite{Specht2009}. 

In \cite{Walborn2003}, a HG pump was used to investigate how the symmetry of the pump beam may affect the symmetry of the down-converted bi-photon states and in turn the outcome of the HOM interference. However, no experiment to date have investigated the HOM effect of entangled down-converted bi-photon states in the HG basis.

Here, we observe the HOM interference of an entangled bi-photon state, produced in spontaneous parametric down-conversion (SPDC) in the HG basis. We use this to demonstrate that the symmetry properties that are valid for HOM interference in the LG basis \cite{Fedrizzi2009} also apply in the case of the HG basis. Being another infinite dimensional basis, the HG modes provide an alternative approach to prepare high-dimensional entangled quantum states for quantum information applications.

\section{Concept}

The LG modes are a good approximation of the Schmidt basis for SPDC states. This implies that each LG mode for one photon is correlated with only one LG mode for the other photon, in a 1 to 1 correspondence. The azimuthal index $\ell$ in the signal photon will only give $-\ell$ in the idler photon and vice versa. In contrast, the HG basis differs significantly from the Schmidt basis for SPDC states. Hence,  the 1-to-1 correspondence is not observed \cite{Walborn2005}. However, the LG and HG modes are related by the following expression \cite{Zauderer1986}
\begin{widetext}
\begin{equation}
(-1)^{p+|\ell|}2^{2p+|\ell|}|p| ! (x\pm \im y)^{|\ell|}L_p^{|\ell|}(x^2+y^2) = \sum_{m=0}^p\sum_{n=0}^{|\ell|}\left(\begin{array}{l}p\\m\end{array}\right)\left(\begin{array}{l}|\ell|\\n\end{array}\right)(\mp \im)^{|\ell|+n}H_{2m+n}(x)H_{2p+|\ell|-2m-n}(y),
\label{LGtoHG}
\end{equation} 
\end{widetext}
where the $\pm$($\mp$) sign on the left (right) of the equal sign corresponds to whether $\ell$ is positive or negative. 

In the LG basis, SPDC produces the state
\begin{equation}
\ket{\Psi} =  \sum_{p,q = 0}^{\infty}\sum_{\ell = 0}^{\infty} \alpha_{p,q;\ell} \ket{\Psi_{p,q;\ell}^+},
\label{SPDC}
\end{equation}
where $\alpha_{p,q;\ell}$ are the complex coefficients in the expansion, and are determined by the pump and crystal parameters. The following two-photon state
\begin{equation}
\ket{\Psi_{p,q;\ell}^+} = \frac{1}{\sqrt{2}} \left(\ket{L_p^\ell}_A\ket{L_q^{-\ell}}_B+\ket{L_p^{-\ell}}_A\ket{L_q^\ell}_B\right) ,
\label{bell}
\end{equation}
is a symmetric Bell-state in the LG basis.  Here $\ket{L_{p,q}^\ell}$ represents a photon in the LG basis with radial index $p$ or $q$ and an OAM value of $\ell$. The subscripts $A$ and $B$ label the photon paths. 

The state in Eq.~(\ref{bell}) can be converted into the HG basis by using Eq.~(\ref{LGtoHG}). This represents a unitary transformation that converts one basis into another. Such local unitary transformations commute with the path-exchange operator, which determines the symmetry properties of the state. As a result, these local unitary transformations do not alter the path exchange symmetry properties of the state. One can define the path-exchange operator by the $2\times 2$ matrix
\begin{equation}
X = \left[ \begin{array}{cc} 0 & {\cal I} \\  {\cal I} & 0 \end{array} \right]
\end{equation}
where ${\cal I}$ is the identity operator. Symmetric and anti-symmetric states are eigenstates of $X$, with eigenvalues $1$ and $-1$, respectively. A change in the basis of the state can be represented by the operation of two identical unitary operations on the respective subsystems of the state $\ket{\psi}\rightarrow\ket{\phi}= (U_A\otimes U_B) \ket{\psi}$, where $U_A=U_B$. Since the path-exchange operator commutes with such local unitary operators $[(U_A\otimes U_B),X]=0$, one can write
\begin{equation}
X\ket{\phi} = X (U_A\otimes U_B) \ket{\psi} = (U_A\otimes U_B) X \ket{\psi} = \pm \ket{\phi} .
\end{equation}
As a result, the symmetry properties of the state is maintained. 

As an example, we show the simple case of converting the symmetric LG state with $p=0$ and $\ell=2$ below 
\begin{align}
	&\ket{L_0^2}_A\ket{L_0^{-2}}_B + \ket{L_0^{-2}}_A\ket{L_0^2}_B \propto 2 \ket{H_{02}}_A \ket{H_{02}}_B\nonumber\\
	&\quad  + 8 \ket{H_{11}}_{A} \ket{H_{11}}_{B} + 2 \ket{H_{20}}_{A}\ket{H_{20}}_{B}\nonumber\\
	&\quad  - 2 \ket{H_{20}}_{A}\ket{H_{02}}_{B} - 2 \ket{H_{02}}_{A}\ket{H_{20}}_{B} .
\label{symconv}
\end{align}

The symmetry is also preserved when decomposing antisymmetric Bell states into the HG basis. Consider, for example the conversion of an antisymmetric Bell state, composed of LG modes with $p=0$ and $|\ell|=2$ 
\begin{align}
	& \ket{L_0^2}_A\ket{L_0^{-2}}_B - \ket{L_0^{-2}}_A\ket{L_0^2}_B \propto\nonumber\\
	&\quad 4 i \left( \ket{H_{02}}_{A}\ket{H_{11}}_{B} - \ket{H_{11}}_{A}\ket{H_{02}}_{B} \right. \nonumber\\
	&\quad \left. + \ket{H_{11}}_{A}\ket{H_{20}}_{B} - \ket{H_{20}}_{A}\ket{H_{11}}_{B} \right) .
\label{asymconv}
\end{align}
A way to determine the symmetry of the state is by using the HOM interferometer, for which a symmetric input two-photon state would always send the two output photons through the same output ports \cite{Walborn2003,lghom}. So if we set the detection to be in the HG basis, we should observe no coincidence events. If the input two-photon state in a HOM interferometer is antisymmetric, then the two photons will always exit through different output ports and only coincidence events will be observed.

\section{Experiment}

\subsection{Experimental Setup}

\begin{figure}[h]
\centerline{\scalebox{1}{\includegraphics[width=0.5\textwidth]{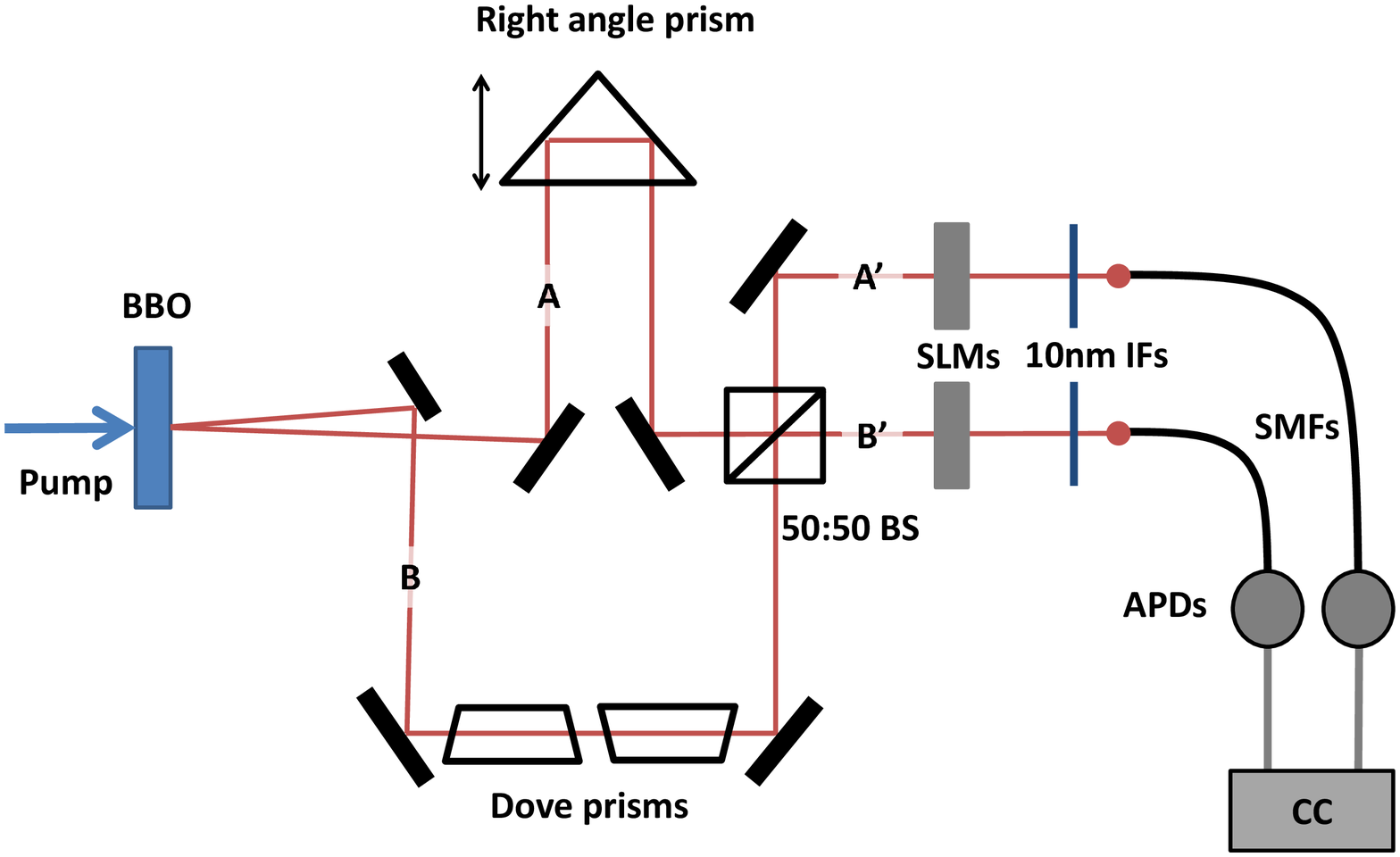}}}
\caption{(Color online) Diagram on experimental setup to observe HOM interference of photons in the HG basis (imaging lenses not shown). BBO - $\beta$-barium borate crystal; APD - Avalanche Photo Diode; CC - Coincidence Counter; SMF - Single Mode Optical Fibre; SLM - Spatial Light Modulator; IF - Interference Filter and BS - non-polarizing 50:50 Beam Splitter.}
\label{setup}
\end{figure}

The experimental setup is shown in Fig.~\ref{setup}. A 350-mW laser with a wavelength of 355 nm is used to pump a 3-mm-thick $\beta$-barium borate (BBO) crystal to produce degenerate photon pairs (labelled $A$ and $B$) with Type-I phase matching. The photons are generated in a non-collinear fashion as this makes them easier to separate by using a D-shaped mirror. The photon in path $A$ is reflected off a right-angle prism mounted on a translational stage that is used to adjust the path length. Path $B$ has two Dove prisms which can be rotated to change the phase of the entangled state \cite{agnew2013}. The photons are then passed through a 50:50 beamsplitter, after which the photons are incident on spatial light modulators (SLMs) encoded with phase only holograms. In combination with single-mode optical fibres (SMFs), these SLMs allow us to make joint projective measurements of particular spatial modes. An interference filter with $\Delta \lambda=10$ nm is used to select out photons around 710 nm just before the SMFs. The SMFs are connected to avalanche photodiodes to detect the single photons and coincidences are registered via a coincidence counter. To maximize the coincidence count rate, we image the photon pairs from the plane of the BBO crystal onto the SLMs and then from the SLMs into the ends of the SMFs. 

\subsection{Experimental Results}

\begin{figure}[h]
\centering
\subfloat[]{
\scalebox{1}{\includegraphics[width=0.5\textwidth]{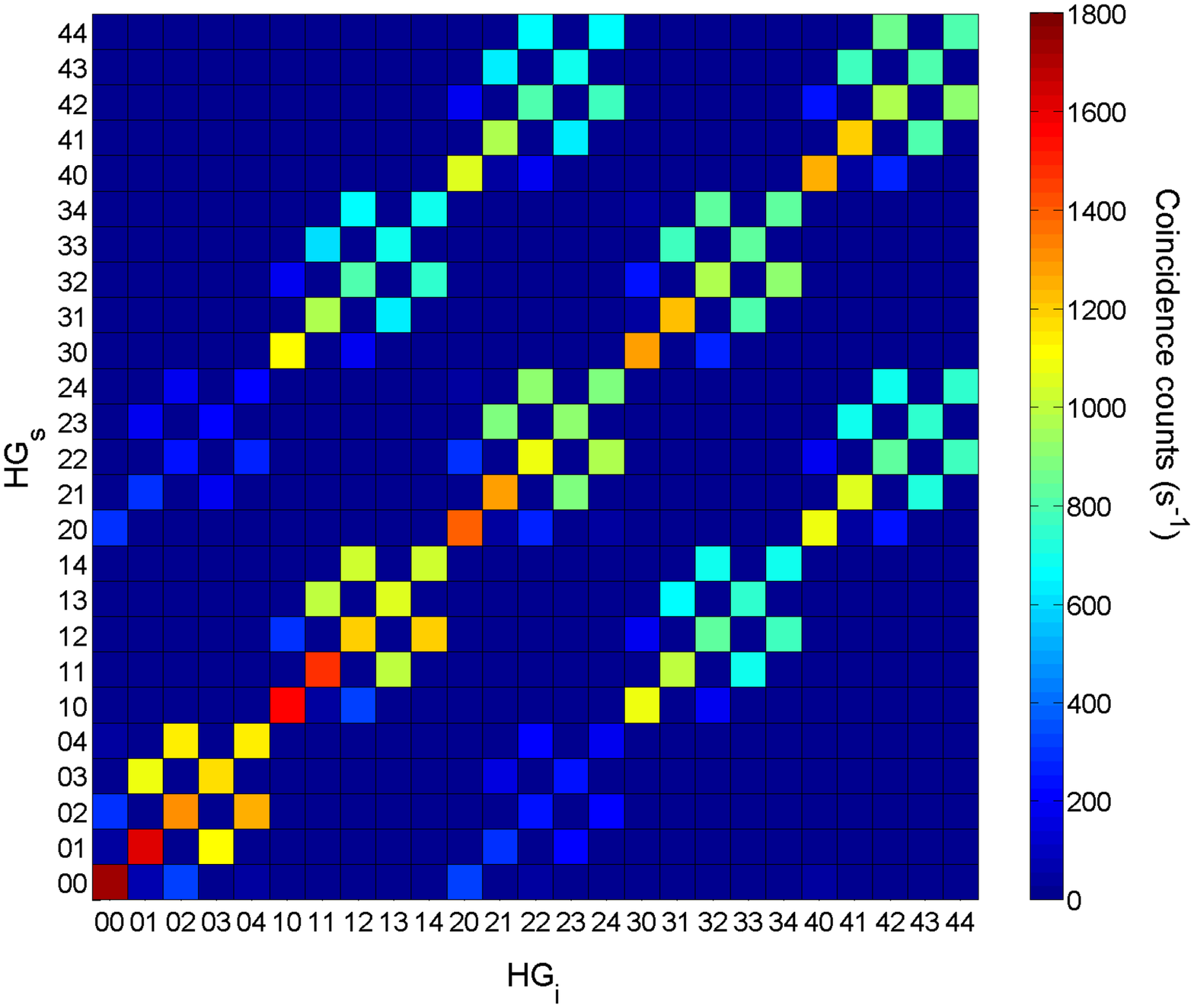}}}\\
\subfloat[]{
\scalebox{1}{\includegraphics[width=0.5\textwidth]{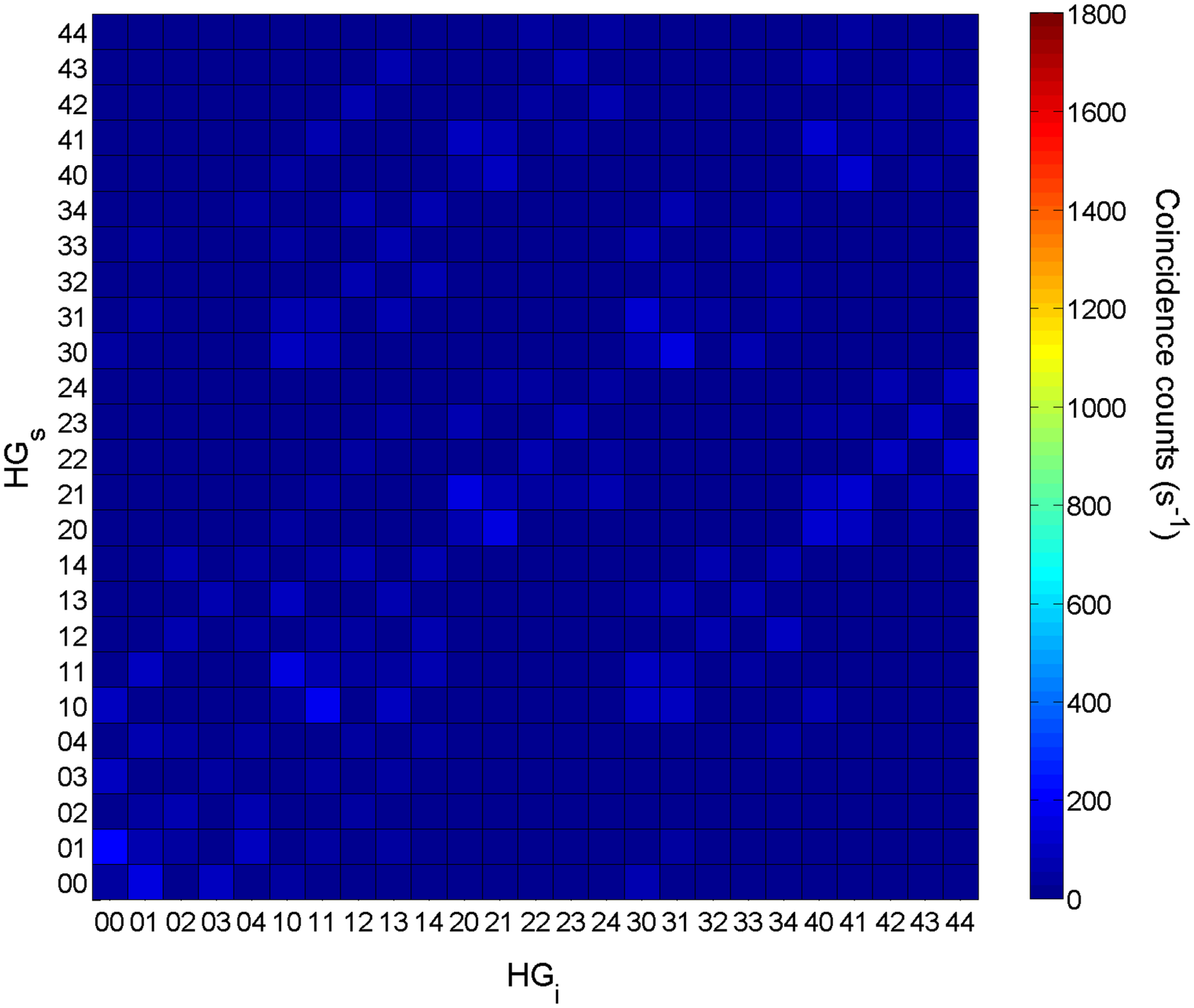}}}
\caption{(Color online) Coincidence counts measured for a spectrum of HG modes when the dove prisms are at $0$ degrees to each other. (a) shows the coincidence counts measured when the path length is unequal and (b) shows the coincidence counts decreases to almost zero for when there is equal path length. }
\label{D0}
\end{figure}

\begin{figure}[h]
\centering
\subfloat[]{
\scalebox{1}{\includegraphics[width=0.5\textwidth]{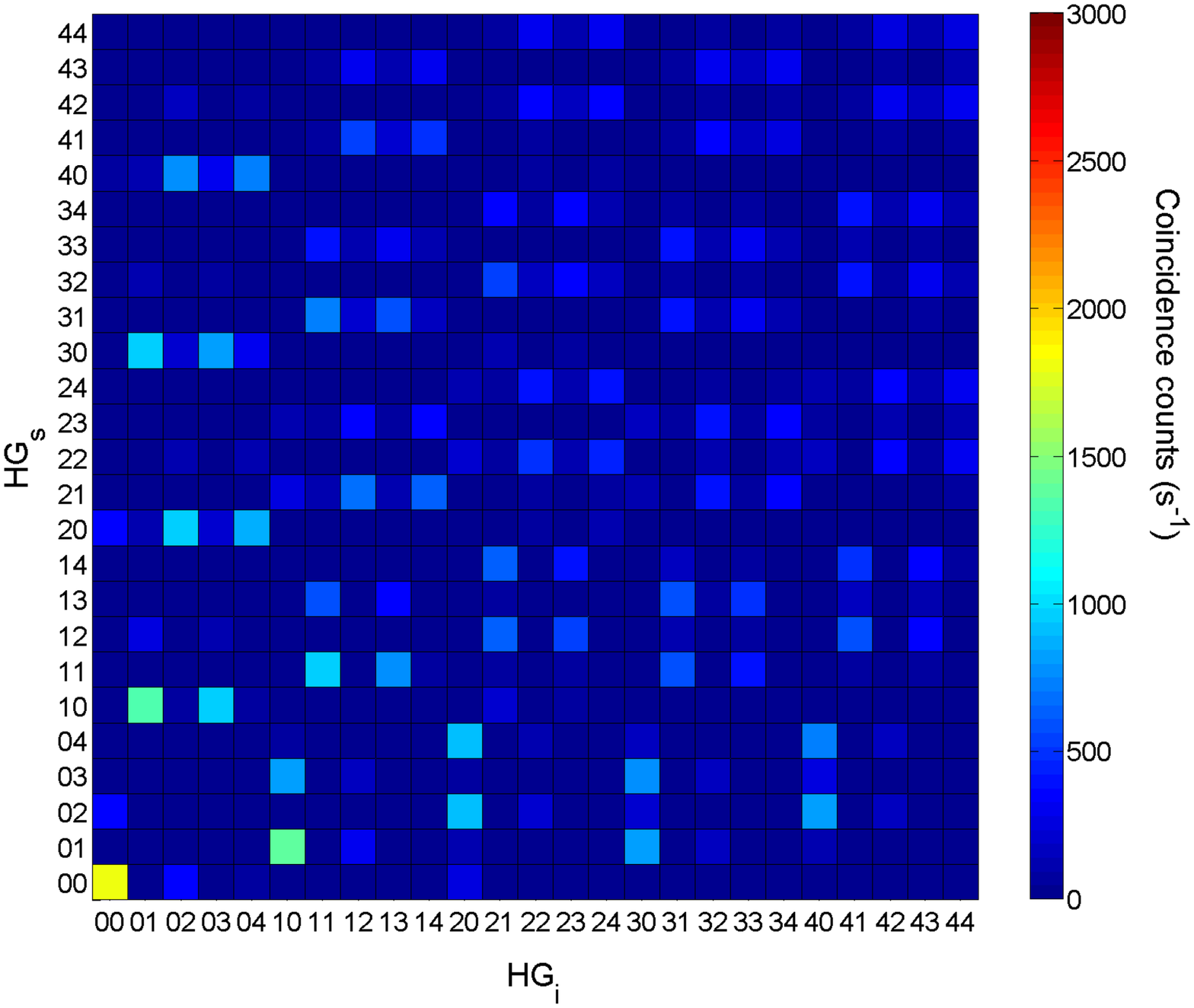}}}\\
\subfloat[]{
\scalebox{1}{\includegraphics[width=0.5\textwidth]{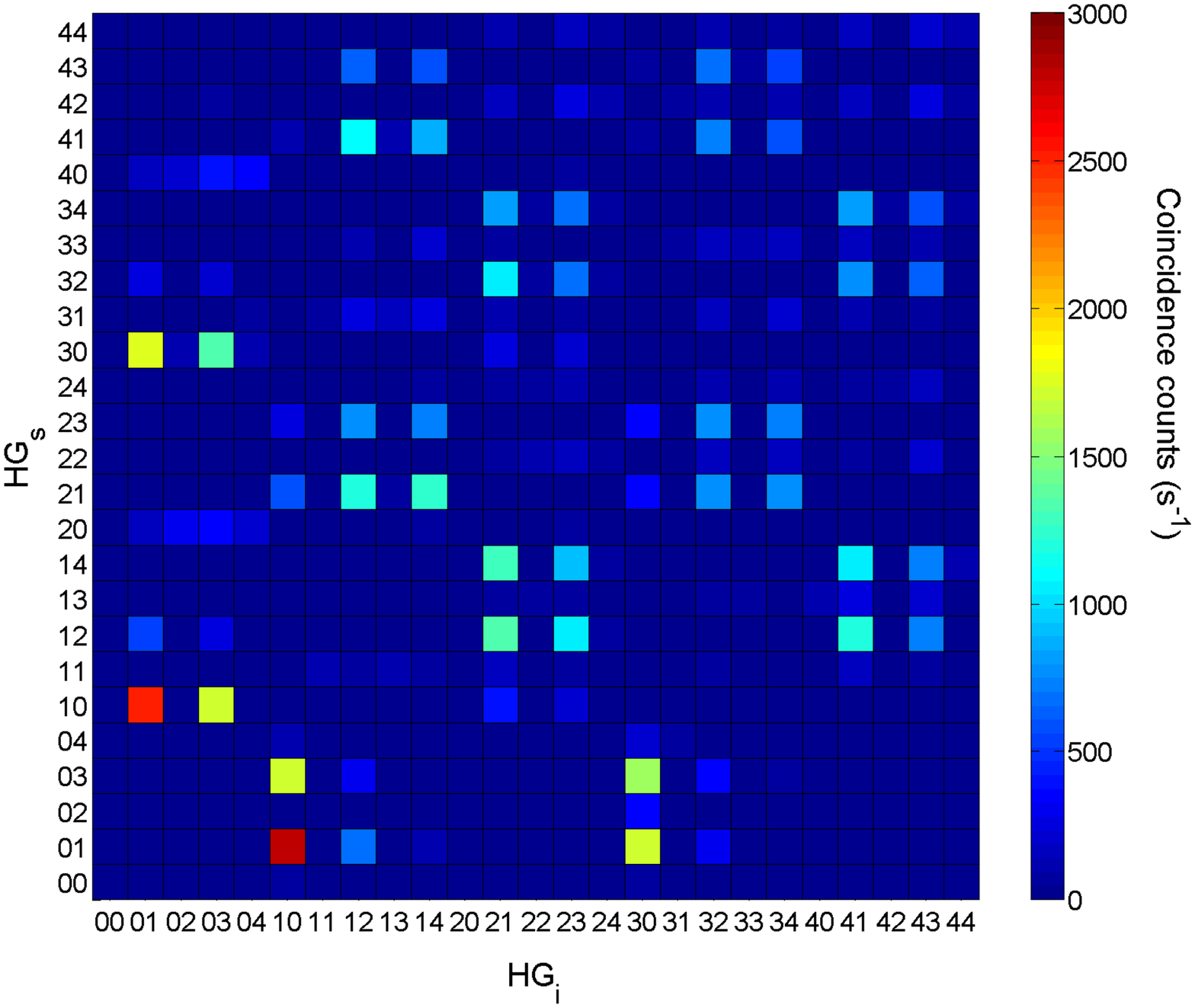}}}
\caption{(Color online) Coincidence counts measured for a spectrum of HG modes when the dove prisms are at $45$ degrees to each other. (a) shows the coincidence counts measured when the path length is unequal and (b) shows that some of the modes have coincidence counts decreased to zero while others doubled when there is equal path length. }
\label{D45}
\end{figure}

In the LG basis, the entangled photon pairs generated through Type-I phase matching are in the symmetric $\Psi^+$ Bell state. When the Dove prisms are set at zero degrees to each other there will be no change in the symmetry properties of the entangled states. If we convert from LG into the HG basis, the symmetry in the state is preserved, as seen in Eq.~(\ref{symconv}). When the input state into a HOM interferometer is symmetric, the two photons will always exit through the same output port, so no coincidence events are observed in the single photon detectors. Setting the two Dove prisms at 45 degrees to each other and measuring in the LG basis, we convert all the $\Psi^+$ states with odd $\ell$'s  into the antisymmetric $\Psi^-$ states, while all the states with even $\ell$'s remain unchanged \cite{lghom}.  Since one only observes coincidence events when a $\Psi^-$ state enters the HOM interferometer,  one only sees coincidence events for HG modes that are decomposed from LG modes with odd $\ell$'s. Those that are decomposed from even $\ell$'s  produce no coincidence events.

In Figs.~\ref{D0} and \ref{D45} we show the numbers of coincidence events for different settings of the Dove prisms. These are measured for all the combinations of HG modes in the signal and idler photons, with indices $m,n$ ranging for 0 to 4. 

The case when the Dove prisms are oriented at zero degrees to each other is shown in Fig.~\ref{D0}. Figure~\ref{D0}(a) shows the coincidence data for the input state without HOM interference and Fig.~\ref{D0}(b) shows the output coincidence data after HOM interference. We see that, as expected, little to no coincidence events are observed in Fig.~\ref{D0}(b) after the HOM interference.

Figure~\ref{D45} shows the case when the Dove prisms are oriented at 45 degrees to each other. The coincidence data for the input state when there is no HOM interference and for the output state when HOM interference is present, are shown in Fig.~\ref{D45}(a) and (b), respectively. It can be seen that, under HOM interference, the coincidence counts of specific combinations of the observed HG modes dropped to zero, while those of the rest are doubled. The sum of the HG indices always add up to $|\ell| + 2p$. In this way, one can see that the combinations of HG modes that produce double the original number of coincidence counts are those related to LG modes with odd $\ell$'s, which become antisymmetric when the Dove prisms are oriented at 45 degrees with respect to each other.

\section{Discussion \& Conclusion}

The HOM interference of spatial modes has been investigated previously \cite{karimi2014,Walborn2003,Nagali2009,Exter2007,Exter2010,lghom}. However, the only one concerned with HG modes \cite{Walborn2003} investigated how the symmetry of a hybrid (multimode) polarization and HG state in the pump beam affects the overall symmetry of the SPDC state and in turn the outcome of the HOM interference. All the other investigations \cite{karimi2014,Nagali2009,Exter2007,Exter2010,lghom} studied the HOM effect in the LG basis.

Here, we investigate the HOM interference for entangled photon pairs in the HG basis. By putting two dove prisms in the path of the photons and rotating them with respect to each other, we were able to manipulate the entangled quantum state which enters the HOM interferometer. We found that if the entangled photon pairs are in a symmetric Bell state in the LG basis, then they will remain symmetric after decomposing into the HG basis, thereby resulting in no coincidence events. If the photon pairs are in an antisymmetric Bell state in the LG basis, then they will also be antisymmetric in the HG basis, thereby producing only coincidence events.

\bibliographystyle{unsrt}
\bibliography{kwant}

\end{document}